\documentclass{scrartcl}

\usepackage{graphicx}
\usepackage{amsmath}
\usepackage{amssymb}
\usepackage[binary-units]{siunitx}
\usepackage{xcolor}

\usepackage{placeins}

\usepackage[pdftex, 
linkcolor=violet,
citecolor=blue,
colorlinks=true,
urlcolor=purple]{hyperref}

\usepackage{caption}
\newcommand\authormark[1]{\textsuperscript{#1}}

\DeclareSIUnit\od{OD}

\begin{document}

\title{Transient absorption microscopy setup with multi-ten-kilohertz shot-to-shot subtraction and discrete Fourier analysis.}

\author{Robert Schwarzl\authormark{1}, Pascal Heim\authormark{1}, Manuela Schiek\authormark{2}, \\
Dario Grimaldi\authormark{3}, Andreas Hohenau\authormark{3}, \\ 
Joachim R. Krenn\authormark{3} and Markus Koch\thanks{corresponding author: markus.koch@tugraz.at}\authormark{1}}

\date{}

\maketitle

\noindent
\authormark{1}Graz University of Technology, Institute of Experimental Physics, Petersgasse 16, 8010 Graz, Austria.\\
\authormark{2}Johannes Kepler University Linz, LIOS \& ZONA, Altenberger Str. 69, A-4040 Linz, Austria.\\
\authormark{3}University of Graz, Institute of Physics, Universitätsplatz 5, 8010 Graz, Austria. \\

\begin{abstract}

Recording of transient absorption microscopy images requires fast detection of minute optical density changes, which is typically achieved with high-repetition-rate laser sources and lock-in detection. 
Here, we present a highly flexible and cost-efficient detection scheme based on a conventional photodiode and an USB oscilloscope with MHz bandwidth, that deviates from the commonly used lock-in scheme and achieves benchmark sensitivity. 
Our scheme combines shot-to-shot evaluation of pump--probe and probe--only measurements, a home-built photodetector circuit optimized for low pulse energies applying low-pass amplification, and a custom evaluation algorithm based on Fourier transformation. Advantages of this approach include abilities to simultaneously monitor multiple frequencies, parallelization of multiple detector channels, and detection of different pulse sequences (e.g., include pump--only). 
With a \SI{40}{\kilo\hertz} repetition-rate laser system powering two non-collinear optical parametric amplifiers for wide tuneability, we demonstrate the 2-D imaging performance of our transient absorption microscope with studies on micro-crystalline molecular thin films.
\end{abstract}

\section{Introduction}

Transient absorption (TA) spectroscopy is a technique where dynamics of a system after photoexcitation are investigated by detecting the intensity change of a probe laser pulse as a function of time, typically with femtosecond resolution~\cite{Megerle2009,Cabanillas-Gonzalez2011,Knowles2018,Lai2020}. With laser beam diameters in the millimeter range, the technique can be used to examine homogeneous samples such as dissolved molecules in solution, amorphous solids and single crystals.
Micro- and nanostructured systems, such as quantum dots, nanowires, or textured thin films,
additionally require spatial resolution to resolve local variations of the investigated processes. In transient absorption microscopy (TAM) \cite{Grumstrup2015a,Fischer2016,Davydova2016a,Zhu2019,Zhu2020,Ahmed2020} the laser pulses are focused to few \si{\micro\meter} or less, often close to their diffraction limit, in order to provide spatial as well as temporal resolution; spatial modulation even allows sub-diffraction-limited resolutions \cite{Massaro2016}.

In TA measurements, the change in optical density is recorded with the pump-probe technique in two consecutive steps. First, a probe pulse passes through the sample where some of its energy is absorbed; the remaining wavelength-dependent intensity $I_\text{pr}(\lambda_\text{pr})$ is then recorded at the detector. Second, femto- to picoseconds before the next probe pulse arrives, a pump pulse excites a certain fraction of the sample (c.f., Fig.~\ref{fig:setup}). This dynamical alteration in state population directly leads to a transient change in probe intensity $I_\text{pu-pr}(\lambda_\text{pu},\lambda_\text{pr}, \Delta t)$, dependent on the pump- and probe wavelength ($\lambda_\text{pu}$ and $\lambda_\text{pr}$ respectively) and on the pump--probe time delay ($\Delta t$),  which is again measured at the detector. The pump beam does not contain relevant information and is therefore discarded. The influence of a pump pulse on the probe pulse is represented by the transient absorbance $\Delta A$, often given in orders of magnitude of optical density (\si{\od}). Eq.~\eqref{eq:transient_absorbance} shows this dependency explicitely:

\begin{equation}
    \Delta A(\lambda_\text{pu},\lambda_\text{pr}, \Delta t) = -\log_{10} \left( \frac{I_\text{pu-pr} (\lambda_\text{pu},\lambda_\text{pr}, \Delta t)}{I_\text{pr} (\lambda_\text{pr})} \right) = -\log_{10} \left( 1 + \frac{\Delta I_\text{pr} (\lambda_\text{pu}, \lambda_\text{pr}, \Delta t)}{I_\text{pr} (\lambda_\text{pr})} \right) \label{eq:transient_absorbance}
\end{equation}

The second half of the equation relates $\Delta A$ to the pump-induced alteration of the probe intensity, $\Delta I_\text{pr}= I_\text{pu-pr}-I_\text{pr}$, which is often stated in literature.

In transient absorption microscopy, the illuminated sample area is typically very small, in consequence of the high spatial resolution required for many samples. In order to avoid photodamage, low fluences are required, often resulting in long data acquisition times. A high sensitivity $\Delta A_\text{min}$ is therefore a fundamental requirement for an efficient TAM setup.
 
Within the past years, multiple approaches have been presented to tackle this problem. Lock-in amplifiers and high repetition rate laser oscillators are often utilized to resolve intensity changes down to $\frac{\Delta I_\text{pr}}{I_\text{pr}} = $\num{1e-7} \cite{Zhu2020} (note that the absolute sensitivity mainly depends on the number of averaged pulses).  Recently, a high repetition rate laser has been combined with a multi-wavelength line scanner to reach $\frac{\Delta I_\text{pr}}{I_\text{pr}}=$\num{1e-6} \cite{Grumstrup2019}. By contrast, amplified laser systems provide a much higher pulse energy at lower repetition rates, which enable the use of optical parametric amplification to generate wavelength-tuneable pulses, or broad-band white light supercontinuua for probing. The sensitivity suffers from the lower repetition rate and lies typically in the range of $ \frac{\Delta I_\text{pr}}{I_\text{pr}} \approx \num{1e-4}$. 

In this publication, we present a detection scheme that deviates from the commonly used direct lock-in analysis of the photodetector signal. We demonstrate the feasibility of using a single photodiode followed by analog signal processing and digitizing at moderate sample rates of \SI{10}{\mega\hertz}. We use a laser repetition rate of \SI{40}{\kilo\hertz} and a mechanical chopper to achieve shot-to-shot acquisition of pump-probe measurements. The digitized signal is analyzed by Fourier transformation. With this approach we achieve a sensitivity of $\frac{\Delta I_\text{pr}}{I_\text{pr}}=\num{4.7e-5}$ for a \num{200000} pulse measurement in \SI{5}{\second} with a beam diameter of \SI{6}{\micro\meter}. With two non-collinear optical parametric amplifiers (NOPAs), we achieve wavelength-tuneability of both pump and probe pulses at a temporal resolution of \SI{80}{\femto \second} (FWHM, cross correlation). For a probe pulse energy above \SI{5}{\pico\joule}, the sensitivity turns out to be almost exclusively limited by the laser noise. This high sensitivity (signal-to-noise ratio) combined with the non-destructive interaction of low energy pulses make the presented setup a versatile tool to study the spatio-temporal properties of a wide range of modern materials, such as two-dimensional transition metal dichalcogenides~\cite{Choi2017}. To demonstrate these capabilities, we investigate dynamics in thin films of micro-structured organic molecular crystals, which are very sensitive to photodamage.

Compared to photodiodes with dedicated lock-in amplifier hardware, the advantages of our approach include (i) the ability to monitor multiple frequency components at once, allowing the determination of pump-probe and probe-only intensities with one detector, (ii) the option to extend the pulse sequence to pump-only background measurements or dump pulses, and (iii) enabling massive parallelization, e.g., for multi-wavelength detection, all without significant extra costs.

\section{Experimental Setup}

Our TAM, sketched in Fig.~\ref{fig:setup}, consists of a femtosecond  pump--probe microscopy setup to achieve femtosecond temporal and micrometer spatial resolution, a home-built photodetector optimized for very low pulse energies, and an analog-to-digital converter (ADC), all three of which are described in the following sections. 
Analysis of the digitized photodetector signal is achieved through digital Fourier transformation, as described in Section~\ref{sec:analysis}.

\subsection{Femtosecond Pump--Probe Microscope}

\begin{figure}[ht]
	\centering
    \includegraphics[width=.9\textwidth]{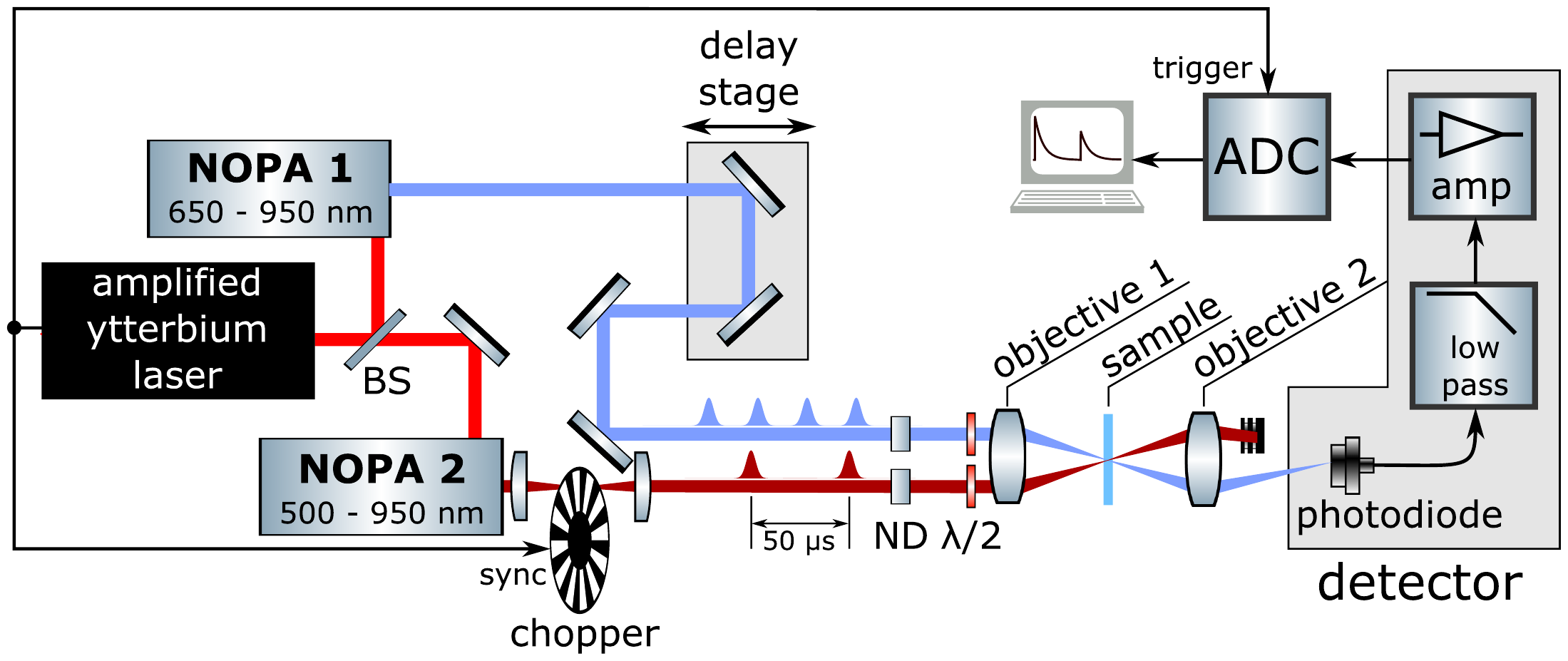}
	\caption{Transient absorption microscope setup. Either of the two NOPAs can be used to generate the pump or probe pulses from the amplified ytterbium laser PHAROS PH1-20. The signal from the home-built photodetector is recorded using a PicoScope 5442A USB oscilloscope (ADC). }
	\label{fig:setup}
\end{figure}

We generate femtosecond laser pulses with an amplified Yb:KGW laser system (Light Conversion PHAROS PH1-20, \SI{400}{\micro\joule} pulse energy, \SI{1025}{\nano\meter} central wavelength, \SI{\sim270}{\femto\second} pulse duration),
operated at \SI{40}{\kilo\hertz} repetition rate. The pulses are evenly split to power two non-collinear optical amplifiers (both from Light Conversion, NOPA 1: ORPHEUS-N-2H, NOPA 2: ORPHEUS-N-3H). The accessible wavelength range stretches from 500/650 to \SI{950}{\nano\meter} (c.f., Fig.~\ref{fig:setup}), with the option for additional frequency doubling. Prism compressors at the output of both NOPAs are used to obtain pulse durations of $\approx$\SI{30}{\femto\second} at the sample.

Because the laser pulse energy fluctuates with a strong correlation of successive laser pulses \cite{Kanal:14,Kearns:17}, which is known as $\frac{1}{f}$ noise, very rapid individual pump-probe measurements increase the signal-to-noise ratio. We therefore use a high-speed mechanical chopper (SciTec 310CD) to perform shot-to-shot measurements by blocking every other pump pulse. 
The pump beam is focused through the \SI{0.76}{\milli\meter} broad slits of the chopping disk (SciTec~300CD200HS) whose rotational speed is synchronized to the pump laser at half the repetition rate. An individual pump-probe event is completed after two successive laser pulses, thus within \SI{50}{\micro\second}, and a measurement averages typically over a few seconds. The pump-probe delay can be set with a delay stage (motorized with a Newport LTA-HS actuator), providing a time-step resolution of \SI{0.67}{\femto\second}. 
The temporal resolution therefore depends almost only on the pulse duration. An intensity cross-correlation measurement yields a temporal resolution of \SI{80}{\femto\second} (FWHM) at \SI{800}{\nano\meter} pump and probe wavelength. For some measurements, reflective or absorptive neutral density filters (ND), irises, achromatic $\lambda/2$ waveplates, or  broadband wire grid polarizers are inserted into the pump and probe paths to adapt intensity and polarizations.

Both beams are spatially overlapped at the sample surface with a first microscope objective (Olympus UPlan FL N 10x, NA$=0.3$), and transmitted light is collected with a second objective (Nikon Plan 10x, NA$=0.3$) in confocal arrangement. The sample is located at the focal plane and can be moved perpendicularly to the optical axis at micrometer precision with motorized translation stages (Thorlabs Z825B actuators). While the transmitted pump beam is blocked with mechanical beam blocks and/or spectral edge-pass filters, the probe light is guided onto the home-built photodetector and digitized, as described in the next section.

\subsection{Photodetector}
\label{sec:detector}

\begin{figure}[h]
	\centering
	\includegraphics[width=.7\textwidth]{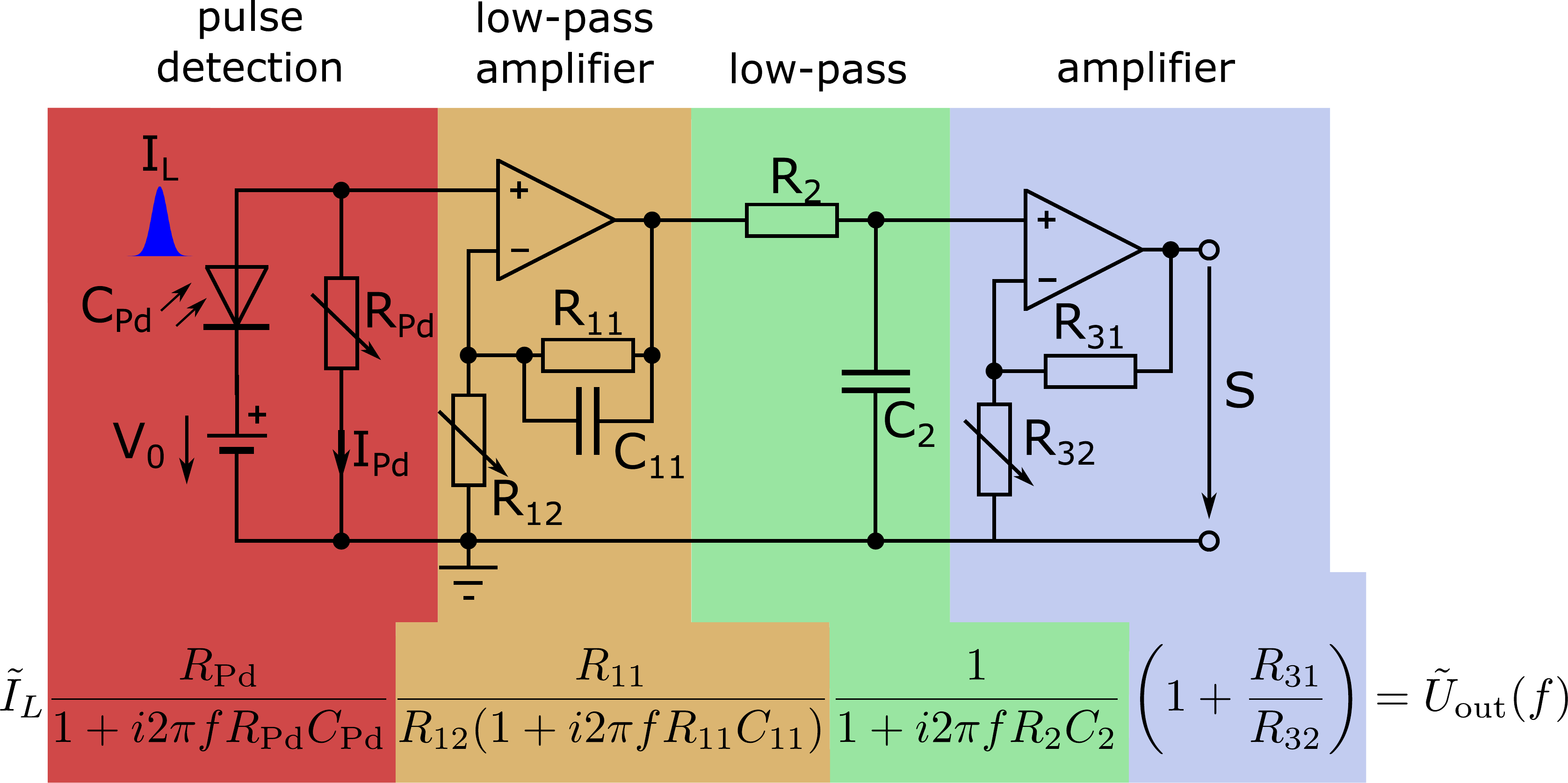}
	\caption[Amplifier circuit schematic]{Amplifier circuit schematic\label{fig:detector_schematic} including the frequency response Eq.~\refstepcounter{equation}(\arabic{equation})\label{eq:frequency-response}. $I_L$ current produced by the light pulse which charges the capacity of the photodiode, $S$ resulting voltage signal. See text for a description of the circuit.}
\end{figure}

In order to detect transmission changes of the probe pulses with the highest possible sensitivity, we use a home-built photodetector, shown in Fig.~\ref{fig:detector_schematic}. The detector is optimized for low laser pulse energies required in TAM. Additionally, it allows for low sample rates in the digitization process of the output signal by stretching the nanosecond voltage pulses from the PD into the \SI{25}{\micro\second} time window between two consecutive laser pulses\cite{Schriever2008}. 
The setup can easily be adapted to different repetition rates by changing the resistor and capacitor values and therefore the frequency response. We have chosen a photodiode (Hamamatsu, S1336-5BQ) with small capacitance ($C_\textrm{Pd} = \SI{65}{\pico\farad}$) in order to account for the low pulse energies in the range of \SI{5}{\pico\joule}, corresponding to only $\approx \num{3.2e7}$
photons per pulse (at \SI{650}{\nano\meter}). With the photo diode's quantum efficiency of \SI{70}{\percent}, the generated voltage pulses peak 
at about \SI{17}{\milli\volt} (for \SI{3}{\pico\joule} pulses), before it decays  with a time constant of $\tau_\textrm{Pd} = C_\textrm{Pd} R_\textrm{Pd}$ ($R_\textrm{Pd}\sim \SI{10}{\kilo\ohm}$).

\begin{figure}[h]
	\centering
  \includegraphics[width=\textwidth]{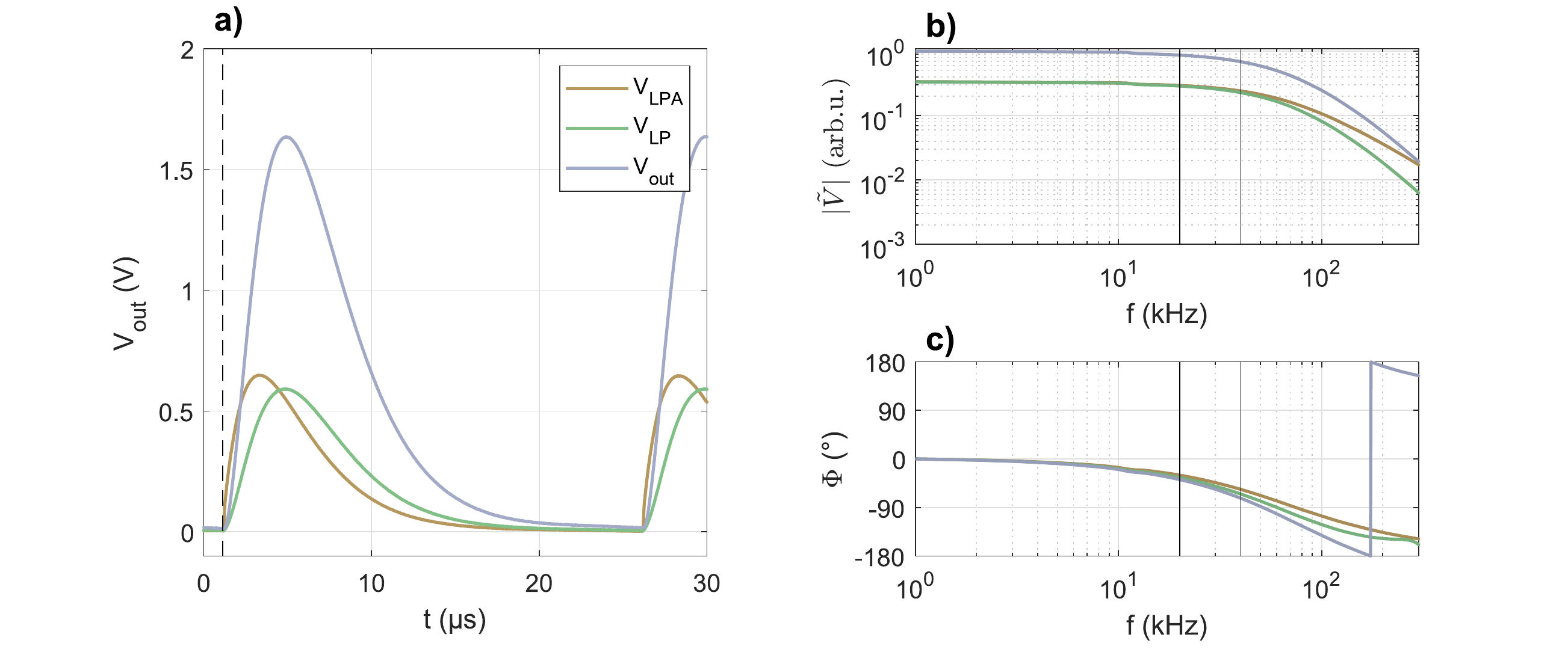}
	\caption{Time- and frequency-domain behavior of the photodetector.
	\textbf{(a)} Voltage pulses at different stages of the amplifier. The dashed line indicates the arrival time of the laser pulse. The voltage at $R_\textrm{Pd}$ is not shown because it cannot easily be measured due to the input impedance of the oscilloscope being too low. The orange and green curves show the stretched pulse shape after the first OP-AMP and after the low pass filter, respectively (see Fig.~\ref{fig:detector_schematic}, similar color coding).  The purple curve is the output voltage and represents the impulse response function $V(t)$ that characterizes the temporal behavior of the photodetector. \textbf{(b), (c)} Bode plot showing the frequency-dependent amplification (b) and phase shift (c) of the amplifier circuit. Vertical lines at 20 and \SI{40}{\kilo\hertz} indicate the chopping frequency and the laser repetition rate, respectively.}
	\label{fig:bode_plot_amplifier}
\end{figure}

The voltage pulse is amplified and further stretched in time with a low-pass amplifier (yellow section in Fig.~\ref{fig:detector_schematic}), using an operational amplifier (OP-AMP) with a capacitor ($C_{11}$) in parallel with the feedback resistor ($R_{11}$). We use an OP-AMP with high input impedance in order to separate the small photodiode current $I_\textrm{Pd}$ from the rest of the circuit. Note that the voltage rise at $R_\textrm{Pd}$ after laser pulse arrival proceeds very rapidly, so that the rise of the OP-AMP output voltage is limited to its slew rate. The duration of this nonlinear behavior lasts a few nanoseconds and is reduced by $C_{11}$, which lowers the amplification factor during this fast voltage change. A measurement of the transient voltage pulse after this active low pass is shown as a red line in Fig.~\ref{fig:bode_plot_amplifier}a. 
The pulse is then further stretched in time by a passive low-pass filter (yellow in Figures~\ref{fig:detector_schematic} and~\ref{fig:bode_plot_amplifier}a), and finally amplified by a second OP-AMP to match the input range of the analog-to-digital converter (purple in Figures~\ref{fig:detector_schematic} and Fig.~\ref{fig:bode_plot_amplifier}a). 

The amplification function in the frequency domain is shown in Fig.~\ref{fig:detector_schematic} (Eq.~\eqref{eq:frequency-response}) with the same color coding as in the schematic, in order to highlight the contributions of the individual amplifier stages. The corresponding Bode plot in Figures~\ref{fig:bode_plot_amplifier}b and 3c shows the frequency response of the individual amplifier stages.
For further analysis of the detected laser pulses in Chapter~\ref{sec:analysis}, we will use the impulse response function $V(t)$ of the output (purple line in Fig.~\ref{fig:bode_plot_amplifier}a) in the time domain, and the frequency response function $\tilde V(f)$ (purple line in Fig.~\ref{fig:bode_plot_amplifier}b) in the frequency domain.

\subsection{Signal Digitization}

The advantage of our photodetector with its low-pass characteristics becomes apparent in the signal digitization stage, where a low-bandwidth and low-sampling-rate ADC suffices. We use a low-cost ADC (Picoscope~5442D) with \SI{10}{\mega\siemens\per\second} sample rate, \SI{60}{\mega\hertz} bandwidth and 16 bit resolution, capable of recording up to \num{128000} pump-probe events in sequence with its on-board memory of \SI{128}{\mega\byte}. 
Due to the low-pass characteristic of our detector, the Nyquist rate is far below the sample rate, so that there is no risk for aliasing.

A typical measurement takes about \SI{1}{\second}, which leads to a frequency resolution (FWHM of $\tilde{S}(\Delta f)$, see below) of \SI{1.2}{\hertz}, further reducing the chance of aliasing problems.

\section{Analysis of Transient Absorption Microscopy Measurements}
\label{sec:analysis}

In the following, we discuss the algorithm for computing the transient absorbance from the digitized signal.
Analysis of the periodic, shot-to-shot TA measurements follows the basic idea of the Lock-In technique \cite{Uhl2021}. 
We modulate the pump pulses at half the laser repetition rate, $f_0/2=\SI{20}{\kilo\hertz}$, by blocking every other pulse (c.f., Fig.~\ref{fig:setup}). This creates a modulation with the same frequency of \SI{20}{\kilo\hertz} in the probe pulse intensity, so that every other probe pulse is increased or decreased in power. 
As the temporal width of the pump pulses is about $10^6$ times shorter than the sampling period of the ADC, the detected laser intensity over time, $I_\textrm{L}(t)$, can be approximated by a delta comb with the alternating amplitudes $I_\textrm{pu-pr}$ and $I_\textrm{pr}$, corresponding to the transmitted probe intensity of pump--probe and the probe-only measurements: 

\begin{align}
    I_\textrm{L}(t) = \sum_{m=-\infty}^{\infty} \Bigg\{I_\textrm{pu-pr} \delta\left[\frac{2}{f_0}m-\Big(t-T_0\Big)\right] + I_\textrm{pr} \delta\left[\frac{2}{f_0}m-\left(t-T_0 - \frac{1}{f_0}\right)\right]\Bigg\},
    \label{eq:ilt}
\end{align}
where $T_0$ is a temporal offset adjustable using the pump laser's internal delay and $m$ is an integer accounting for the signal periodicity.
This periodic laser signal is detected with the photodetector, consisting of the photodiode and low-pass amplifier (see Fig.~\ref{fig:detector_schematic}), and subsequently digitized. Thus the signal $S(t)$ can be described as convolution of the laser pulses $I_\textrm{L}(t)$ with the impulse response function of the detector $V(t)$ (see Fig.~\ref{fig:bode_plot_amplifier}):

\begin{align}
    S(t) = I_\textrm{L}(t) \circledast V(t)
    \label{eq:S_equals_I_conv_V}
\end{align}
In the frequency domain, the signal spectrum $\tilde{S}(f)$ is obtained as product of the pulse spectrum  $I_\textrm{L}(f)$ with the frequency response function  $\tilde{V}(f)$ 
\begin{align}
    \tilde{S}(f) = \tilde{I}_\textrm{L}(f) \tilde{V}(f).
\end{align}
Due to the alternating intensity of the laser pulses arriving at the detector (Eq.~\ref{eq:ilt}), corresponding to a $1/(2f_0)$ periodicity, the signal spectrum is discrete with a $f_0/2$ spacing (see Eq.~\ref{eq:app_fourier_laser_pulses} in the supplemental information).
\begin{align}
    \tilde{S}\left(n\frac{f_0}{2}\right) = e^{i\phi_0 \frac{n}{2}} \tilde{V}\left(n\frac{f_0}{2}\right) \left[|\tilde{I}_\textrm{pu-pr}| + (-1)^n |\tilde{I}_\textrm{pr}| \right]
    \label{eq:S_tilde_of_I}
\end{align}
where $n$ is an integer, $\phi_0 = 2\pi f_0 T_0$ corresponds to the temporal offset $T_0$. $|\tilde{I}_\textrm{pu-pr}|$ and $|\tilde{I}_\textrm{pr}|$ are proportional to the alternating amplitudes $I_\textrm{pu-pr}$ and $I_\textrm{pr}$, which we can determine from the two equations for $n=1~ \textrm{and}~2$:

\begin{align}
n&=1:    \tilde{S}\left(\frac{f_0}{2}\right) = e^{i \frac{\phi_0}{2}} \tilde{V}\left(\frac{f_0}{2}\right) \left(|\tilde{I}_\textrm{pu-pr}| - |\tilde{I}_\textrm{pr}| \right)
\label{eq:Sf2}\\
n&=2:    \tilde{S}\left(f_0\right) = e^{i\phi_0} \tilde{V}\left(f_0\right) \left(|\tilde{I}_\textrm{pu-pr}| + |\tilde{I}_\textrm{pr}| \right)
\label{eq:Sf}
\end{align}

$\left|\tilde{S}\left(\frac{f_0}{2}\right)\right|$ thus is proportional to the difference of two consecutive pulses, which is typically very small, while $\left|\tilde{S}\left(f_0\right)\right|$ is proportional to their sum.
Since the detector influences both amplitude and phase in dependence of frequency, it is important to consider the frequency response function $\tilde{V}(f)$ for the two frequencies $f_0$ and $\frac{f_0}{2}$ (c.f., Fig.~\ref{fig:bode_plot_amplifier}).

We define two detector-specific quantities. First, $P_{f_0}$, which has unit length and accounts for the phase shift induced by the detector at $f_0$:
\begin{align}
 P_{f_0} &= \frac{\tilde{V}(f_0)^*}{|\tilde{V}(f_0)|} \label{eq:pf0-calculation}
\end{align}

Second, $T_{\frac{f_0}{2}\rightarrow f_0}$ accounts for both gain difference and phase shift difference between $f_0$ and $\frac{f_0}{2}$:
\begin{align}
   T_{\frac{f_0}{2} \rightarrow f_0} &= \frac{\tilde{V}(f_0)}{\tilde{V}(\frac{f_0}{2})} \label{eq:tf02f0-calculation}
\end{align}
Both complex quantities, $P_{f_0}$ and $T_{\frac{f_0}{2} \rightarrow f_0}$, are determined from the experiment and are constant for stable detector configurations (see section~\ref{sec:detector-property-measurement} in the supplemental information.)
\\

With these quantities equations~(\ref{eq:Sf2}) and (\ref{eq:Sf}) can be written as:
\begin{align}
\tilde{S}\left(\frac{f_0}{2}\right) T_{\frac{f_0}{2} \rightarrow f_0} P_{f_0} e^{-i \frac{\phi_0}{2}}
&= \left|\tilde{V}\left(f_0\right)\right| \left(|\tilde{I}_\textrm{pu-pr}| - |\tilde{I}_\textrm{pr}| \right) =: \mathcal{S}\left(\frac{f_0}{2}\right) 
\label{eq:ssf2}\\
\tilde{S}\left(f_0\right) P_{f_0} e^{-i\phi_0} 
&=\left|\tilde{V}\left(f_0\right)\right| \left(|\tilde{I}_\textrm{pu-pr}| + |\tilde{I}_\textrm{pr}| \right) =:  \mathcal{S}(f_0) 
\label{eq:ssf}
\end{align}

Note that $\mathcal{S}\left(\frac{f_0}{2}\right)$ and $\mathcal{S}(f_0)$ are a real-valued quantities because the phase shift induced by the detector is compensated by $ T_{\frac{f_0}{2} \rightarrow f_0}$ and $P_{f_0}$  and the phase induced by the time difference between the first pulse and $t=0$ is compensated by $\phi_0$. 
The phase shift $\phi_0$ can be calculated in every measurement separately from
\begin{align}
 \tilde{S}(f_0)P_{f_0} &= |\tilde{S}(f_0)|e^{i\phi_0} 
\end{align}
as
\begin{align}
 \phi_0 &= \arccos \left\{ \frac{\textrm{Re} \left[ \tilde{S}(f_0)P_{f_0} \right]}{\left| \tilde{S} (f_0) \right|} \right\}
\end{align}

The transient absorbance change $\Delta A$, as defined in Eq.~(\ref{eq:transient_absorbance}), can thus be calculated from the frequency-domain signals $\tilde{S}\left(\frac{f_0}{2}\right)$ and $\tilde{S}\left(f_0\right)$ with Equations~(\ref{eq:ssf2}) and (\ref{eq:ssf}):

\begin{align}
    \Delta A = -\log_{10}\left(\frac{I_\textrm{pu-pr}}{I_\textrm{pu}}\right)
    = 
    -\log_{10}\left(\frac{|\tilde{I}_\textrm{pu-pr}|}{|\tilde{I}_\textrm{pu}|} \right)
    =
    -\log_{10}\left\{\frac{\textrm{Re}\left[\mathcal{S}(f_0) + \mathcal{S}\left(\frac{f_0}{2}\right)\right]}
    {\textrm{Re}\left[\mathcal{S}(f_0) - \mathcal{S}\left(\frac{f_0}{2}\right)\right]}\right\}
\end{align}

Note that $\mathcal{S}(f_0)$ and $\mathcal{S}\left(\frac{f_0}{2}\right)$ in Eq.~\eqref{eq:ssf2} and \eqref{eq:ssf} has no imaginary part, but noise has. Therefore, by discarding the imaginary part, the signal-to-noise ratio is significantly improved.
\\
The evaluation uses the phase and amplitude of certain frequencies. If a signal is sampled discretely in time, then the phase of the Fourier transform is very sensitive on the frequency $f$. In fact, a frequency change of $f \rightarrow f + \frac{1}{\Delta T}$, where $\Delta T$ is the sample time of the oscilloscope, leads to a phase shift of $2\pi$. Therefore, it is crucial to evaluate the repetition rate $f_0$ in each measurement in order to compensate for slight drifts caused by external influences on the laser system. A method for determining repetition rate drifts is explained in section~\ref{ap:reprate} of the supplemental information.

Section~\ref{sec:appendix-multiple} of the supplemental information shows the extension of the data analysis method for arbitrarily many delta combs. This would allow measuring the pump-only background by modulating the pump and the probe pulses in order to make a pump-probe, pump-only and probe-only measurement.

\section{Performance of the TAM Setup}

\subsection{Dynamics in Thin Metal Films: Sensitivity Determination}
\label{sec:sensitivity}

\begin{figure}[h]
	\centering
    \includegraphics[width=.9\textwidth]{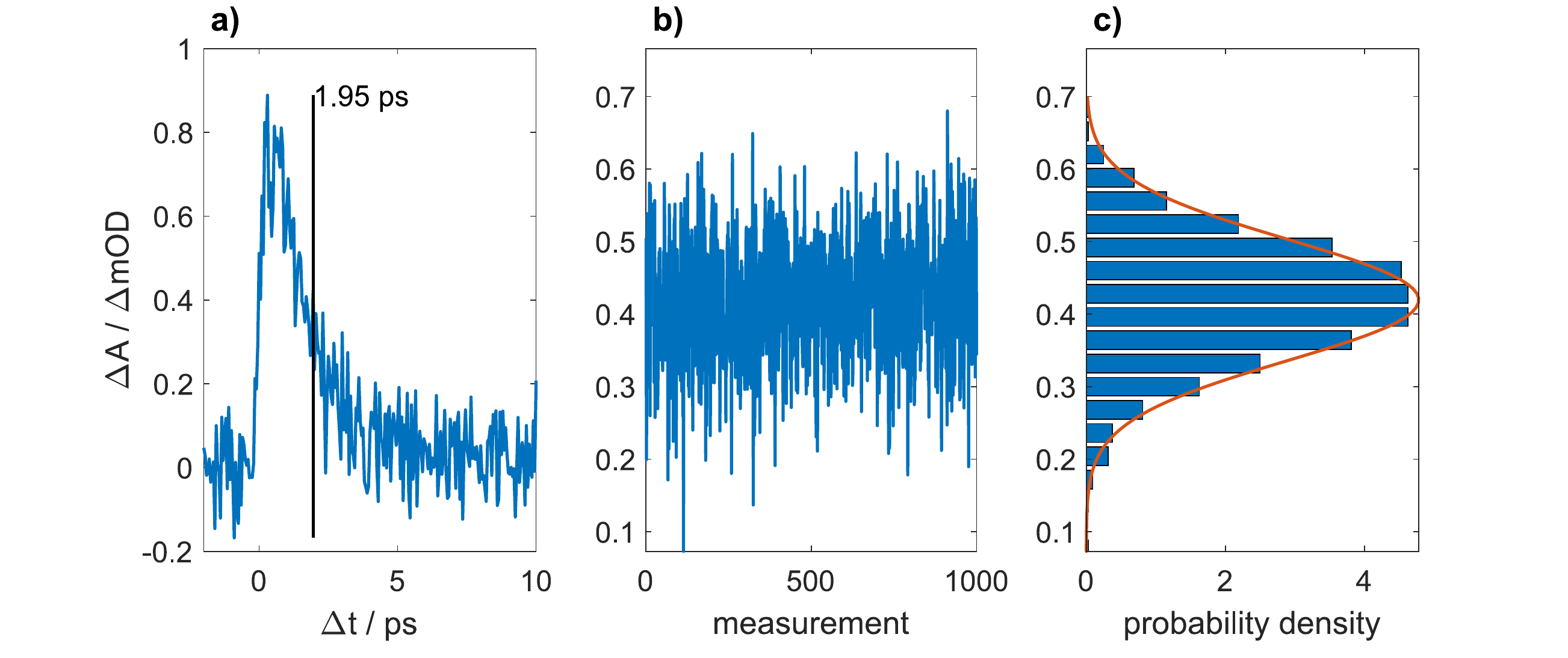}
	\caption{Transient absorption measurement series of a thin metal film (integration time \SI{5}{\second} per point) for sensitivity determination. 
	\textbf{(a)} Pump-probe delay scan. 
	\textbf{(b)} Dataset of 1000 measurements at a constant pump-probe delay of \SI{1.95}{\pico\second}, with a data acquisition time of \SI{0.5}{\second} for each measurement.
	\textbf{(c)} Probability distribution of the data set shown in (b), following a Gaussian distribution with a standard deviation of \SI{65}{\micro\od} corresponding to $\frac{\Delta I_\text{pr}}{I_\text{pr}}=$\num{1.5e-4}.
	}
	\label{fig:sensitivity_measurement}
\end{figure}

We determine the sensitivity of our setup by measuring the picosecond dynamics of thin metal films. We use a thin film of \SI{30}{\nano\meter} gold and \SI{3}{\nano\meter} chromium deposited on a commercial ITO-coated microscopy slide (indium tin oxide). Similar measurements which also probe the transition from the d-band to the Fermi surface on gold thin films can be found in the literature~\cite{PhysRevB.86.155139}. They show that the contributing factors to the transient absorption signal are the thermalization of electrons through electron-electron interaction and electron-phonon interaction; with dominance of the latter on the \si{\pico\second} timescale. Fig.~\ref{fig:sensitivity_measurement}a shows the transient absorption obtained with pump pulses of \SI{680}{\nano\meter} wavelength (\SI{1.82}{\electronvolt} photon energy), \SI{275}{\pico\joule} pulse energy and \SI{\approx 1050}{\micro\joule\per\square\centi\meter} peak fluence, and probe pulses of \SI{504}{\nano\meter} (\SI{2.46}{\electronvolt}), \SI{30}{\pico\joule} and \SI{\approx 115}{\pico\joule\per\square\centi\meter}. Pulses of approximately \SI{10}{\pico\joule} are transmitted through the sample and recorded by the detector. The laser spot diameter at the sample was \SI{6}{\micro\meter}. In order to determine the signal fluctuations, we record a dataset of \num{1000} individual measurements at a pump-probe delay of \SI{1.95}{\pico\second}, each of which accumulates data for \SI{0.5}{\second} at \SI{40}{\kilo\hertz} repetition rate, corresponding to \num{10000} probe-only and \num{10000} pump-probe events. Fig.~\ref{fig:sensitivity_measurement}b shows the individual transient absorption values and Fig.~\ref{fig:sensitivity_measurement}c a histogram of their relative abundance, which follows a normal distribution (red curve), as expected. The standard deviation of the data set is $\sigma (\Delta A) = \SI{65}{\micro\od}$ corresponding to $\frac{\Delta I_\text{pr}}{I_\text{pr}}=$~\num{1.5e-4} for a single \SI{0.5}{\second} measurement. Note that for a typical high-resolution measurement with an integration time of \SI{5}{\second} (\num{200000} laser pulses), the standard error reduces to \SI{20}{\micro\od} corresponding to $\frac{\Delta I_\text{pr}}{I_\text{pr}}=$~\num{4.7e-5}.

In order to identify the origin of these fluctuations, we compare the dataset to a simulation, explained in detail in section~\ref{sec:appendix-noise} of the supplemental information. This simulation is based on the pulse energy distribution collected with our photodetector by integrating over single pulses from the home-built photodetector at a repetition rate of \SI{20}{\kilo\hertz}. The reduced repetition rate avoids overlapping pulses in the digitized signal. The pulse energy distribution is in agreement with measurements generated from a Coherent LabMax pulse meter at \SI{400}{\hertz}.
Based on the assumption of independent pump and probe pulses, the simulation predicts a  larger standard deviation of  $\sigma (\Delta A) = \SI{111}{\micro\od}$ corresponding to $\frac{\Delta I_\text{pr}}{I_\text{pr}}=$~\num{2.6e-4}. This result indicates a pulse energy correlation of consecutive laser pulses, leading to reduced fluctuations in the shot-to-shot analysis. Importantly, this result also indicates that the sensitivity of our TAM is limited by pulse energy fluctuations of the NOPAs, and that the contribution to the fluctuations of our photodetector and evaluation algorithm are negligible. We note that we typically obtain lower fluctuations around the center of the NOPA tuning curve, where the device performance is more stable.

\subsection{Exciton Dynamics In Micro-Crystalline Molecular Thin Films}

We demonstrate the performance of our TAM with the investigation of micro-textured organic thin films. With a film thickness of \SI{\sim50}{\nano\meter} resulting in a weak transient absorption, these molecular crystals require a high sensitivity. 
The sample under investigation consists of anilino squaraines with isobutyl side chains (SQIB) in its orthorhombic crystal structure. The samples are obtained on glass substrates by solution processing with subsequent thermal annealing to induce crystallization into platelet-like rotational domains with preferred parallel orientation of a single crystallographic plane to the substrate.~\cite{Balzer2017,Funke2021} The SQIB platelet size of about \SI{100}{\micro\meter} and their characteristic linear polarized absobance pattern are well suited for microscopic optical investigations. The SQIB platelets are characterized by a pronounced Davydov splitting of the excited states into a lower Davydov component (LDC) at \SI{1.68}{\electronvolt} (\SI{740}{\nano\meter} photoexcitation wavelength) and an upper Davydov component (UDC) at \SI{1.91}{\electronvolt} (\SI{650}{\nano\meter}) \cite{Balzer2017}. The two Davydov components can be selectively excited due to this pronounced splitting, and because the transition dipole moments of the UDC and LDC are perpendicular, (for a full characterization of the biaxial dielectric tensor with imaging Mueller matrix ellipsometry see Ref.~\cite{Funke2021}). This results in a characteristic linear dichroism with two absorption maxima (UDC and LDC) polarized mutually perpendicular within the plane of a platelet.
Since the platelets have a random rotational in-plane orientation, transmission of linearly polarized light with fixed wavelength and polarization direction results in different spatial absorbance patterns of each platelet. \cite{Funke2021}

\begin{figure}[ht]
    \centering
    \includegraphics[width=.9\textwidth]{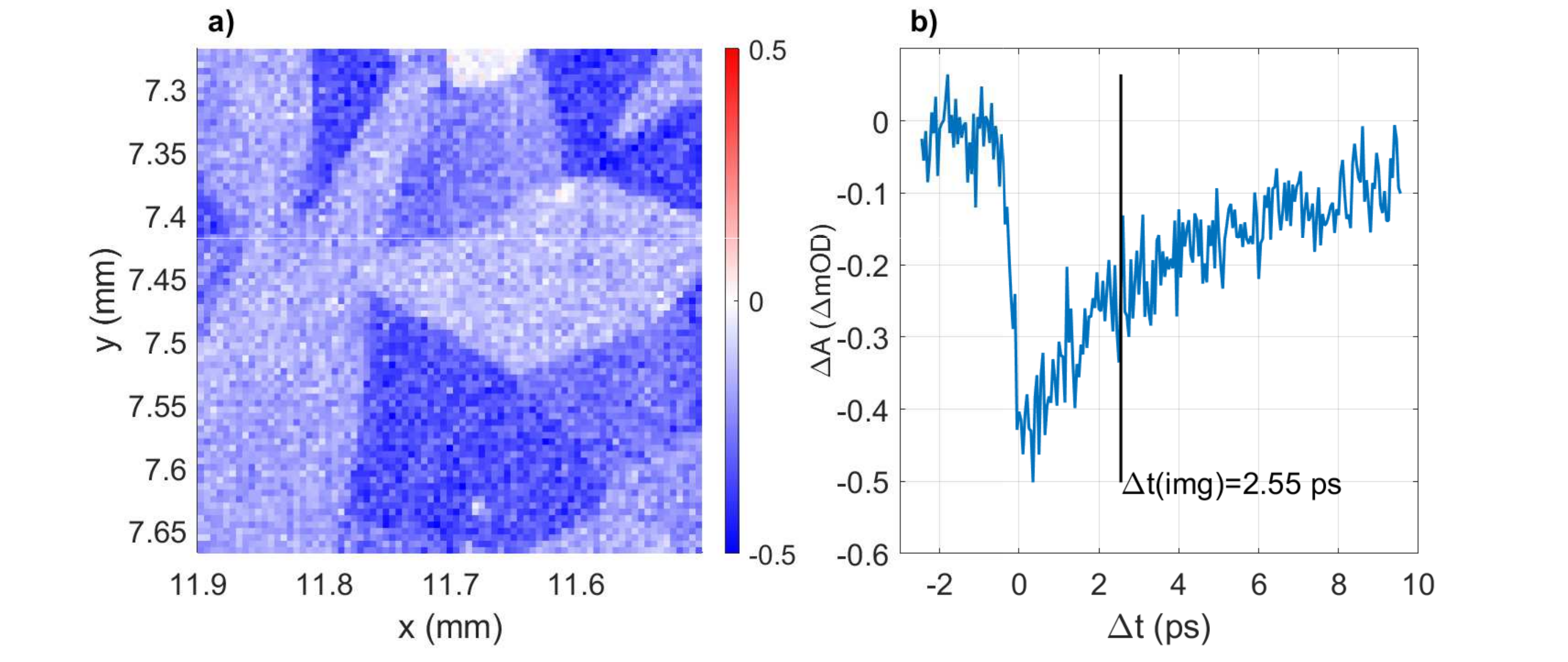}
    \caption{TAM measurements of micro-textured orthorhombic SQIB, obtained by UDC excitation and time-delayed probing of the LDC band. Pump pulses  (\SI{12.5}{\pico\joule}) and probe pulses (\SI{6.3}{\pico\joule}) are perpendicularly polarized.   
    \textbf{(a)} TAM image recorded at \SI{2.55}{\pico\second} pump-probe delay with spatially fixed polarizations.
    \textbf{(b)} TA scan of a single platelet with optimized pump and probe polarization directions.
  For the image in (a), the transient absorbance $\Delta A$ without pump influence is subtracted from the pump-probe transient absorbance at \SI{2.55}{\pico\second} to remove stray pump light. The FWHM of the pump and probe beam is \SI{4}{\micro\meter}.
    }
    \label{fig:ta-imaging-sqib}
\end{figure}

Here, we demonstrate that our TAM setup is capable to observe the ultrafast exciton dynamics within individual platelets. Fig.~\ref{fig:ta-imaging-sqib} shows a TAM measurement with pump excitation to the UDC and probe absorption at the transition from the ground state to the LDC. We rotate the polarization direction of the pump beam in order to obtain maximum single-pulse absorbance in the chosen platelet. The probe beam is polarized perpendicularly to the pump. 
The transient absorption shown in Fig.~\ref{fig:ta-imaging-sqib}b becomes instantly negative, then increases exponentially within about \SI{10}{\pico\second} and finally levels off  at a slightly negative value. This transient behavior is indicative of immediate ground-state bleach followed by rapid and almost complete non-radiative population decay to the ground state, which is in agreement with the low fluorescence yield observed in static experiments~\cite{Chen1994}.  

A 2D image recorded with a fixed time delay of \SI{2.55}{\pico\second} is shown in Fig.~\ref{fig:ta-imaging-sqib}a. The individual platelets become visible, because the transient absorption signal $\Delta A$ depends on the relative orientation of the laser polarization and the UDC transition dipole moment of the platelets. This measurement indicates that the setup is also capable of polarization-resolved femtosecond microscopy by introducing $\frac{\lambda}{2}$ plates and polarizing optical elements.

\section{Conclusion and Outlook}

We present a highly sensitive, flexible and yet cost-efficient detection scheme for transient absorption microscopy. A sensitivity of $\frac{\Delta I_\text{pr}}{I_\text{pr}}=$~\num{4.7e-5} corresponding to $\Delta A=$~\SI{20}{\micro\od} is achieved using a Fourier transform based analysis algorithm. This algorithm is applied to alternating pump-probe and probe-only pulses recorded with a home-built photodetection system. The photodiode signal is amplified, low-pass filtered and recorded using a USB oscilloscope. Analysis of the digitized signal could alternatively be achieved with a software-based lock-in amplifier code~\cite{Uhl2021}. Measurements indicate that the dominant source of noise is the femtosecond laser system, so further improvement in signal-to-noise ratio is possible with a more stable laser. Intensities as low as \SI{10}{\pico\joule} per pulse are sufficient to achieve this signal-to-noise ratio. 
\\
This setup allows us to investigate samples that are very sensitive to photodamage such as the textured squaraine organic thin films shown in Fig.~\ref{fig:ta-imaging-sqib}. In the future, we plan to investigate population dynamics and higher excited states of prototypical micro-crystalline textured organic thin films from e.g. squaraines which are not accessible from the ground state.
The detection algorithm allows to record arbitrary pulse trains from one or more channels at the same time. This is useful for pump-dump-probe setups, separate pump-only measurements and more. Furthermore, a spatially separated pump-probe extension will provide insight into charge or energy propagation on the \si{\micro\meter} scale.


\section*{Acknowledgements}
PH, RS and MK acknowledge the financial support by Zukunftsfonds Steiermark, NAWI Graz, and  the Austrian Science Fund (FWF) under Grant P 33166.
MS thanks the Linz Institute of Technology (LIT-2019-7-INC-313 SEAMBIOF) for funding.

\appendix

\section{Supplemental Material}
\setcounter{equation}{0}
\renewcommand{\theequation}{S.\arabic{equation}}

\subsection{Fourier Transformation of the Signal}
\label{ap:fourier}
As femtosecond laser pulses are \num{1e-6} times shorter than the sampling time of \SI{\approx 100}{\nano\second}, they can be approximated as a periodic delta comb:
\begin{align}
    I_\textrm{L}(t) = \sum_{m=-\infty}^{\infty} \Bigg\{I_\textrm{pu-pr} \delta\left[\frac{2}{f_0}m-\Big(t-T_0\Big)\right] + I_\textrm{pr} \delta\left[\frac{2}{f_0}m-\left(t-T_0 - \frac{1}{f_0}\right)\right]\Bigg\}
\end{align}
Performing a Fourier transform results in a delta comb in the frequency domain:

\begin{align}
    \tilde{I}_L(f) 
    &= 
    \int_{-\infty}^{\infty} I_\textrm{L}(t) e^{-i2\pi f t} dt \notag \\
    &=
    e^{-i2\pi T_0 f} \big[I_\textrm{pu-pr} + I_\textrm{pr}e^{-i2\pi \frac{f}{f_0}} \big] \underbrace{\sum_{m=-\infty}^{\infty} e^{-i2\pi f \frac{2}{f_0}m}}_{= \sum_{n=-\infty}^{\infty}\delta \left( f-n\frac{f_0}{2} \right)} \notag \\
    &=
    \sum_{n=-\infty}^{\infty} e^{-i\pi T_0 f_0 n} \big[I_\textrm{pu-pr} + (-1)^nI_\textrm{pr} \big] \delta \left( f-n\frac{f_0}{2} \right)
    \label{eq:app_fourier_laser_pulses}
\end{align}
where $n$ and $m$ are integers accounting for the periodic nature of the combs.

\subsection{Repetition Rate Measurement}
\label{ap:reprate}
The repetition rate $f_0$ is estimated after every measurement to account for slow repetition rate drifts. An efficient way to estimate the repetition rate $f_0$ is to use a frequency $f$ which is near $f_0$. Then, a sliding window Fourier transformation is performed at this frequency with a rectangular window:
\begin{align}
    \tilde{S}(f,\tau) &= \frac{1}{N}\sum_{j=1}^{N} S(t_j+\tau) e^{-i2\pi f (t_j+\tau)} \notag \\ 
    &=
    \sum_{n=-\infty}^\infty \tilde{A}\left(\frac{f_0}{K}n\right)\tilde{I}\left(\frac{f_0}{K}n\right)e^{i\pi \left(\frac{f_0}{K}n-f\right) (T + 2\tau)} \frac{\sin\left(\pi \left(\frac{f_0}{K}n-f\right) \Delta t N\right)}{N \sin\left(\pi \left(\frac{f_0}{K}n-f\right) \Delta t\right)},
\end{align}
where $K$ is an integer resulting from the Fourier transformation of the Dirac comb. By using the fact that the window size $T = (N-1)\Delta t \gg \frac{1}{f_0}$ we can approximate the above equation with only the term where $\frac{f_0}{K}n - f$ is minimal. Let us assume that is the case for $n=K$ where $K$ is the number of involved delta combs:
\begin{align}
    \tilde{S}(f,\tau)
    \approx
    \tilde{A}\left(f_0\right)\tilde{I}\left(f_0\right)e^{i\pi (f_0-f) (T + 2\tau)} \frac{\sin(\pi (f_0-f) \Delta t N)}{N \sin(\pi (f_0-f) \Delta t)} = |\tilde{S}(f,\tau)|e^{i\Phi(f,\tau)}
\end{align}
    The phase $\Phi(f,\tau)$ of the above expression depends linearly on $\tau$, where the slope is proportional to the frequency offset $f_0-f$:
\begin{align}
    \Phi(f, \tau) = 2 \pi (f_0-f) \tau + \textrm{const.} \label{eq:f0_determination}
\end{align}
Therefore, the repetition rate can be efficiently updated after each measurement.\\

\subsection{Detector property measurement}
\label{sec:detector-property-measurement}
First, the laser is set to $\frac{f_0}{2}=\SI{20}{\kilo\hertz}$ and only pump or probe pulses are recorded. Therefore, the recorded signal is approximately a delta comb at $\frac{f_0}{2}$ and uniform peak height. The detector properties $P_{f_0}$ and $T_{\frac{f_0}{2}\rightarrow{}f_0}$ are then determined using the following algorithm:
\begin{enumerate}
    \item Perform a single beam measurement at $\frac{f_0}{2}$, simultaneously recording the voltage across the photodiode resistor $V_\text{Pd}(t_i)$ and the signal voltage $V_\text{out}(t_i)$ where $i$ is the sample number.
    \item Calculate $f_0$ from the signal $V_\text{out}$ according to \eqref{eq:f0_determination}.
    \item Determine temporal delay $\Delta t$ between trigger and pulse at $V_\text{Pd}$ via slope detection: $t_0=t(V_\text{i+1,Pd}-V_\text{i,Pd} > \Delta V_\text{threshold} )$ (the threshold is set manually by checking if the pulse is detected correctly on the timeline).
    \item Perform discrete fourier transform at $\tilde{I'} (f_0)=\mathcal{F} \{V_\text{out}\} (f_0)$ and $\tilde{I'} \left( \frac{f_0}{2} \right)=\mathcal{F} \{V_\text{out}\} \left( \frac{f_0}{2} \right)$.
    \item Correct $\tilde{I'}(f_0)$ and $\tilde{I'} \left( \frac{f_0}{2} \right)$ for the phase introduced by the pulse delay $\Delta t$, a phase introduced before the detector:
    \begin{align*}
        \tilde{I} (f_0) = \tilde{I'}(f_0) e^{2 \pi i f_0 \Delta t} \\
        \tilde{I} \left( \frac{f_0}{2} \right) = \tilde{I'} \left( \frac{f_0}{2} \right) e^{\pi i f_0 \Delta t}
    \end{align*}
\end{enumerate}

This effectively sets $I_\text{pu-pr}$ in Eq.~(8) and (9) in the manuscript to $I_\text{pr}$, allowing for the calculation of Eq.~(10) and (11) from the timeline:
\begin{align}
    T_{\frac{f_0}{2}\rightarrow{}f_0} &= \frac{\tilde{I} (f_0)}{\tilde{I} \left( \frac{f_0}{2} \right) } \\
    P_{f_0} &= \frac{\tilde{I} (f_0)^*}{\left| \tilde{I} (f_0) \right|}
\end{align}

\subsection{Arbitrary Delta Combs}
\label{sec:appendix-multiple}
The energy $I(t)$ deposited at the detector can be approximated as several delta combs $I_k(t)$ similar to sec.~\ref{ap:fourier}, where $K$ represents the number of different involved delta combs, $\frac{1}{f_0}$ is the time difference between two delta combs and $T_0$ is an temporal offset.: 
\begin{align}
    I_k(t)
    &\approx
    I_k \sum_{m=-\infty}^\infty \delta\left(\frac{K}{f_0}m - (t-T_0)\right) \\
    &= 
    |\tilde{I}_k| \sum_{n=-\infty}^{\infty} e^{i\phi_0 \frac{n}{K}} e^{i2\pi n\frac{f_0}{K}(t-\frac{k-1}{f_0})}
    \\
    I(t)
    &=
    \sum_{k=1}^K I_k(t)
\end{align}

Note that $\phi_0 = -2\pi f_0 T_0$ and that the pulse which arrives between $t=-\frac{T_0}{2}$ and $t=\frac{T_0}{2}$ has the index $k=1$. Because we can approximate $I_k$ as a delta comb, $|\tilde{I}_k\left(n\frac{f_0}{K}\right)|$ is constant.\\

The first few harmonics of $\frac{f_0}{K}$ reveal a set of linear equations:
\begin{align}
    \mathcal{F}(I)\left(n\frac{f_0}{K}\right) = \underbrace{ \left[e^{i\phi_0n} \sum_{k=1}^K|\tilde{I}_k|~ e^{-i2\pi n\frac{f_0}{K}\frac{k-1}{f_0}}\right]}_{\tilde{I}\left(n\frac{f_0}{K}\right)} e^{i2\pi n\frac{f_0}{K}t}
\end{align}

The signal $S$ measured at the A-D converter is a convolution of the laser pulse intensity $I$ and the detector and amplification circuit response $A$ times a constant (which we set to 1):
\begin{align}
    S(t) &= A\circledast I = \mathcal{F}^{-1}(\mathcal{F}(A) \mathcal{F}(I))
\end{align}
This convolution is a product in frequency space:
\begin{align}
    \tilde{S}(f) &= \mathcal{F}(S) = \mathcal{F}(A) \mathcal{F}(I)
\end{align}

In order to obtain the different intensities $I_k$ the frequency dependent amplification of the detection system $\tilde{A}(f)$ at certain frequencies ($\frac{f_0}{K}$, $2\frac{f_0}{K}$, .., $f_0$) must be known. \\
Therefore the detector specific parameters are defined as:
\begin{align}
    P_{f_0} &= \frac{\tilde{A}(f_0)^*}{|\tilde{A}(f_0)|} \\
    T_{n\frac{f_0}{K} \rightarrow f_0} &= \frac{\tilde{A}(f_0)}{\tilde{A}(n\frac{f_0}{K})}
\end{align}
These complex values describe the induced phase shift and amplitude gain induced by the detector and amplification circuit for certain frequencies. These parameters are measured only once for the detector.

Using the Fourier transformed signal $\tilde{S}$ at the first $K$ harmonics of $\frac{f_0}{K}$ and the detector specific parameters a set of linear equations can be established:
\begin{align}
    \mathcal{S}(f_0) &= \tilde{S}(f_0)P_{f_0} = \left[\sum_{k=1}^K |\tilde{A}(f_0)\tilde{I}_k|\right] e^{i\phi_0}\\
    \phi_0 &= \arctan(\mathcal{S}(f_0)) \\
    \mathcal{S}\left(n\frac{f_0}{K}\right) 
    &= 
    \tilde{S}\left(n\frac{f_0}{K}\right)T_{n\frac{f_0}{K}\rightarrow f_0}P_{f_0}e^{-i\phi_0 \frac{n}{K}} \notag \\
    &=
    \sum_{k=1}^K |\tilde{A}(f_0)\tilde{I}_k| e^{-i2\pi n\frac{(k-1)}{K}}
\end{align}
This linear set of equations can easily be solved for $|\tilde{A}(f_0) \tilde{I}_k|$. $\tilde{S}(f_0)$ has a high signal-to-noise ratio and can therefore be used to measure the time delay $T_0$ (by measuring $\phi_0$).

\subsection{Noise Estimation}
\label{sec:appendix-noise}

In order to estimate the influence of the shot-to-shot fluctuations on the determined change in optical density simulations were performed. In a single estimation $N$ laser pulse intensities are drawn from a Gaussian distribution fitted to experimental laser pulse intensity data. This data is generated by measuring the pulse energy distribution of both NOPAs using the photodetector method described in Sec.~4.1 of the manuscript. We assume that the change in optical density $\Delta A$ depends linearly on the pump intensity:
\begin{align}
    \Delta A = \frac{d \Delta A}{d I_\textrm{pump}} I_\textrm{pump}
\end{align}
and that all laser pulses are uncorrelated. $\mathcal{S}_1^{(j)}$ and $\mathcal{S}_1^{(j)}$, which are also measured in a single experiment using $N$ laser pulses, are then computed with the following formula:
\begin{align}
    \mathcal{S}_1^{(j)} &= \frac{1}{N} \sum_{j=1}^N \left( 10^{-\frac{d\Delta \textrm{OD}}{dI_\textrm{pump}}  I_\textrm{pump}^{(1,j)}} I_\textrm{probe}^{(1,j)} + \beta I_\textrm{pump}^{(2,j)} + I_\textrm{probe}^{(2,j)} \right) \\
    \mathcal{S}_2^{(j)} &= \frac{1}{N} \sum_{j=1}^N \left( 10^{-\frac{d\Delta \textrm{OD}}{dI_\textrm{pump}}  I_\textrm{pump}^{(1,j)}} I_\textrm{probe}^{(1,j)} + \beta I_\textrm{pump}^{(2,j)} - I_\textrm{probe}^{(2,j)} \right)
\end{align}
Subsequently, estimating these parameters allows us later to compute a distribution of the corresponding change in optical absorbance $A$.\\

The index $i$ in $I_\textrm{pump}^{(i,j)}$ separates the pump (probe) intensities in probe only ($i=2)$ and pump-probe measurements ($i=1$). The factor $\beta$ is used to model an eventual pump background signal in a pump-probe measurement.

$\Delta A^{(j)}$ of a single measurement can then be estimated by using:
\begin{align}
    \Delta A^{(j)} &= -\log_{10}\left( \frac{\mathcal{S}_1^{(j)}+\mathcal{S}_2^{(j)}}{\mathcal{S}_1^{(j)}-\mathcal{S}_2^{(j)}}\right)
\end{align}

Then the standard deviation of $\Delta A^{(j)}$ can be computed in order to estimate the laser intensity induced variation on the measured $\Delta A$.

\end{document}